\input harvmac
\input epsf.tex
\overfullrule=0mm
\newcount\figno
\figno=0
\def\fig#1#2#3{
\par\begingroup\parindent=0pt\leftskip=1cm\rightskip=1cm\parindent=0pt
\baselineskip=11pt
\global\advance\figno by 1
\midinsert
\epsfxsize=#3
\centerline{\epsfbox{#2}}
{\bf Fig. \the\figno:} #1\par
\endinsert\endgroup\par
}
\def\figlabel#1{\xdef#1{\the\figno}}
\def\encadremath#1{\vbox{\hrule\hbox{\vrule\kern8pt\vbox{\kern8pt
\hbox{$\displaystyle #1$}\kern8pt}
\kern8pt\vrule}\hrule}}

%%%%%%%%%%%%%%%%%%%%%%%%%%%%%%%%%%%%%%%%%%%%%%%%%%%%%%%%%%%%%%%%%%%%%%%%%%%%
% References 
\lref\mac{I.G. Macdonald, S\'eminaire Lotharingien, Publ. I.R.M.A.
Strasbourg, 1988 .}
\lref\Haldcorr{F.D.M. Haldane, M.R. Zirnbauer, Phys. Rev. Lett. Vol. 71,
No. 24, 4055.}
\lref\hal{F.D.M. Haldane, To appear in "Proceedings of the 16th
Taniguchi Symposium", 1993.}
\lref\alti{B.D. Simons, P.A. Lee and B.L. Altshuler, Nucl. Phys. B409,
(1993) 487-508.}
\lref\simon{B.D. Simons, P.A. Lee, and B.L. Altshuler, 
Phys. Rev. Lett. V.70, No26, (1993) 4122.}
\lref\calo{
F. Calogero, J. Math. Phys. {\bf 10}, 2191, (1969).}
\lref\su{B. Sutherland, Phys.Rev.B 38 (1988), 6689.} 
\lref\Haldconj{F.D.M. Haldane, To appear in proceedings of
the International Colloquium in Modern Field Theory, Tata 
institute, 1994.}
\lref\Forres{P. J. Forrester, Nucl. Phys. B388 (1992), 671-699.}
\lref\tables{E.R. Hanson, {\it A Table of Series and Products},
Prentice Hall, N.J., 1975.}
\lref\stan{R. P. Stanley, Adv. in Math., 77, (1989) 76-115.}
\lref\Haldspin{F.D.M. Haldane, Phys.Rev.Lett. 66 (1991) 1529.}
\lref\Haldstat{F.D.M. Haldane, Phys.Rev.Lett. 67 (1991) 937.}
\lref\fad{L.D.Faddeev Les Houches lectures, Elsevier Science Publishers (1984).}
\lref\macdolivre{I.G. Macdonald, {\it Symmetric functions and Hall
polynomials}, Clarendon Press, (1979).}
\lref\Ha{Z.N.C.Ha, F.D.M.Haldane, Phys.Rev.D 47 (1993) 12459.}
\lref\nous{D.Bernard, M.Gaudin, F.D.M.Haldane and V.Pasquier, J.Phys.A 26
(1993) 5219.}
\lref\gaud{M.Gaudin, "la fonction d'onde de Bethe" Masson (1981).}
\lref\drin{V.G.Drinfeld, Funct.Anal.Appl.20 (1988) 56.}
\lref\fel{V.M Buchstaber, G.Felder and A.P. Veselov preprint march 94.}
\lref\cherl{I.V.Cherednick, preprint march 94.}
\lref\cherll{I.V.Cherednick, Commun.Math.Phys (1992) 150.}
\lref\cher{I.V.Cherednick, Invent.Math.106 (1991) 411.}
\lref\nouss{F.Lesage, V.Pasquier and D.Serban saclay preprint april 94.}
\lref\didina{D.Bernard, V.Pasquier and D.Serban saclay preprint april 94.}
\lref\ha{F.D.M Haldane, Phys.Rev.Lett. 60 (1988) 635.}
\lref\sha{B.S.Shastry, Phys.Rev.Lett. 60 (1988) 639.}
\lref\ino{V.I.Inotzemtsev, J.Stat.Phys.59 (1990) 1143.}
\lref\Dunkl{C.F.Dunkl, Trans.Amer.Math.Soc, 311 (1989) 167.}
\lref\yang{C.N.Yang Phys.rev.Letters, 19 (1967) 1312.}
\lref\poli{A.P.Polychronakos,Phys.Rev.Letters, 69 (1992) 703.}
\lref\polil{J. Minanan and A.P. Polychronackos CERN preprint april 94. }
%%%%%%%%%%%%%%%%%%%%%%%%%%%%%%%%%%%%%%%%%%%%%%%%%%%%%%%%%%%%%%%%%%%%%%%%%%%%

\Title{SPhT/94-060}
{{\vbox {
%\centerline{}
\bigskip
\centerline{A lecture on the}
\centerline{Calogero-Sutherland models.} }}}
\bigskip
\centerline{V. Pasquier }

\bigskip

\centerline{ \it Service de Physique Th\'eorique de Saclay
\footnote*{Laboratoire de la Direction des Sciences 
de la Mati\`ere du Commissariat \`a l'Energie Atomique.},}
\centerline{ \it F-91191 Gif sur Yvette Cedex, France}

\vskip .5in

In these lectures, I review some recent results
on the Calogero-Sutherland model and the Haldane Shatry-chain.
The list of topics I cover are the following:
1) The Calogero-Sutherland Hamiltonien and fractional
stastistics. The form factor of the density operator.
2) The Dunkl operators and their relations with 
monodromy matrices, Yangians and
affine-Hecke algebras.
3) The Haldane-Shastry chain in connection with the
Calogero-Sutherland 
Hamiltonian at a specific coupling constant.

\noindent
\vskip 2.0in
%{\noindent \it Submitted to Nuclear Physics B.}
\Date{05/94}

\newsec{INTRODUCTION}

%In these lectures, I review some work done in collaboration 
%with D. Bernard, M. Gaudin, D. Haldane and more recently with F. Lesage and 
%D. Serban on the Calogero-Sutherland model and the Haldane Shatry-chain.
%
%Since there are already several review talks on the subject, see for example
%\hal , I shall 
%only concentrate on a few topics  
%putting forward some technical tools which have 
%proved to be useful in the study of these models.
%
%The list of topics I shall review are the following:
%
%1)\nobreak\ The Calogero-Sutherland Hamiltonian and it's
%connection 
%with fractional statistics.
%The form factor of the density operator.
%
%2)\nobreak\ The Dunkl operators and their relations with 
%monodromy matrices, Yangians and
%affine-Hecke algebras.
%
%3)\nobreak\ The Haldane-Shastry chain in connection with the
%Calogero-Sutherland 
%Hamiltonian at a specific coupling constant.

The Calogero-Sutherland model has recently attracted some attention
mainly because, in spite of its simplicity, it yields nontrivial results which
contribute to shape our understanding of fractional statistics \hal.
Its most remarkable property is that its ground state is given by
a Jastrow wave-function which is the one dimensional analogue of the
Laughlin wave-function. The excitations are described by quasiparticles
which carry a fraction of the quantum numbers of the fundamental particles.
The wave functions are simple enough to allow the computation of
physical quantities such as the dynamical correlation functions.
Recently, it has become clear that the spin version of this model
is closely related to the known spin chains such as the XXX chain.
The wave functions of the spin models can be obtained as simply as those
of the Calogero-Sutherland model by diagonalising simple differential
operators known as the Dunkl operators \Dunkl.

In these lectures, I restrict to the simplest models and I introduce some
techniques useful in their study. These lectures are based on work
done in collaboration with D.Bernard, M.Gaudin and D.Haldane \nous~
and with F.Lesage and D.Serban \nouss.

In the first part, I review Sutherlands method to 
diagonalise the Calogero-Sutherland Hamiltonian \su. I then
give the expression of the form factor of the density operator.
On this example, I describe the fractional character of the quasiparticles
which propagate in the intermediate states.

In the second part, I review the spin generalisation of the Calogero-Sutherland
model and I diagonalise their Hamiltonian using the Dunkl operators.
I then exhibit a representation of the Yangian which commutes with
the Hamiltonian. This is achieved by ``quantizing'' the spectral
parameters of a monodromy matrix obeying the Yang-Baxter equation.
The consistancy of this procedure requires that the
spectral parameters (the Dunkl operators) obey the defining relations
of a affine-Hecke-algebra. Finally, I obtain a representation of this
algebra which degenerates to the Dunkl operators using some operators
defined by Yang in his study of the $\delta$-interacting gas \yang.

In the last part, I review the long range interacting spin model
known as the Haldane-Shastry chain \ha \sha.
Although closely related to the Calogero-Sutherland model, this chain
is more difficult to study because the Dunkl operators cannot be
used to obtain the wave-functions in a straightforward way.
I exhibit a representation of the affine Hecke algebra in terms
of parameters. This representation becomes reducible at the special point where it
commutes with the Haldane-Shastry Hamiltonian.
I then use a correspondance with the Calogero-Sutherland models
to obtain the eigenvalues and some of the eigenvectors of this Hamiltonian.
 
The most unphysical feature of the models 
discussed here is the long range character
of the particle particle interaction which behaves as $1/ x^2$
in the thermodynamical limit.
More realistic models with short range potential $(1/\sinh (x)^2)$
can be defined \ino~ but are more difficult to study (see \cherl~
for some recent progress).
Some aspects of the conformal limit 
of the Haldane-Shastry chain are also discussed in \didina.

\newsec{THE CALOGERO-SUTHERLAND HAMILTONIAN}

This model describes particles on a circle interacting with a long range 
potential $\lbrack$3,4$\rbrack$. The positions of the particles are denoted by
$ x_i, $ $ 1\leq i\leq N, $ $ 0\leq x_i\leq L $ 
and the total momentum and Hamiltonian which give their dynamics are 
respectively given by:
\eqn\hami{
\eqalign{ P & = \sum^ N_{j=1}{1 \over i}{ {\rm d} \over {\rm d} x_j} 
\cr H & =- \sum^ N_{j=1}{1 \over 2}{ {\rm d}^ {\rm 2} \over {\rm d}
x^2_j}+\beta (\beta -1){\pi^ 2 \over L^2} \sum^{ }_{ i<j}{1 \over {\rm sin}^2
\left({\pi \over L} \left(x_i-x_j \right) \right)}  \cr}} 
From now on, we shall work on the unit
circle and set $ \theta_ j=2\pi{ x_j \over L} $ and $ z_j= {\rm e}^{i\theta_
j}. $ 
The wave functions solution of the equation $ H\psi =E\psi $ can be given the
following 
structure:
\eqn\onde{
 \psi (\theta )=\phi (\theta )\Delta^{\beta} (\theta ) }
with
\eqn\del{ \Delta (\theta )= \prod^{ }_{ i<j} {\rm sin} \left({\theta_ i-\theta_ j
\over 2} \right) }
$ \Phi $ is a symmetric polynomial in the variables $ z_j= {\rm e}^{i\theta_j}
$ and $ z^{-1}_j. $

To understand the simplicity of the spectrum, one must write the 
effective Hamiltonian $ \tilde H=\Delta^{ -\beta} H\Delta^{ \beta} $ 
acting on $
\phi . $

\noindent Following Sutherland \su ,  we obtain:
\eqn\sut{
 \eqalign{\tilde H & = \sum^ N_{j=1} \left(z_j{\partial \over \partial
z_j} \right)^2+\beta \sum^{ }_{ i\not= j}{z_i+z_j \over z_i-z_j}
\left(z_i{\partial \over \partial z_i}-z_j{\partial \over \partial z_j}
\right)   \cr}} 
The remarkable property of this Hamiltonian is that acting upon symmetric 
polynomials of a given homogeneity in the variables $ z_j $ it is realised as
a 
triangular matrix. So, it's eigenvalues can be read on the diagonal of the 
matrix and there are simple algorithms to find the eigenvectors.

More precisely, let us define the following basis
of symmetric polynomials 
indexed by a partition: $ \lambda_ 1\geq \lambda_ 2...\geq \lambda_ N\geq 0. $
\eqn\mla{ m_{\{ \lambda\}} = \sum^{ }_{ } z^\lambda }
which is the sum over distinct permutations of the monomial $ z^{\lambda_
1}_1z^{\lambda_ 2}_2...z^{\lambda_ N}_N. $

First, it is easy to see that the subspace of polynomials of a given 
homogeneity $ (\mid \lambda \mid = \sum^ N_{i=1}\lambda_ i) $ is preserved.
Then, inside this subspace, we can define an order on 
the partition by saying that $
\lambda \geq \mu $ 
if $ \lambda_ 1\geq \mu_ 1, $ $ \lambda_ 1+\lambda_ 2\geq \mu_ 1+\mu_
2,...,\lambda_ 1+...+\lambda_ N\geq \mu_ 1+\mu_ 2+...+\mu_ N. $ 
It follows from
Sutherland's 
argument that:
\eqn\tri{ \tilde Hm_\lambda = \sum^{ }_{ \mu \leq \lambda} C_{\lambda \mu} m_\mu
}
that is to say $ \tilde H $ is a triangular matrix. The eigenvalues are
given by the 
diagonal elements:
\eqn\diag{ E_\lambda =C_{\lambda \lambda} = \sum^ N_{i=1} \left(\lambda_ i+(N-i)\beta
\right)^2-((N-i)\beta)^ 2 }
The momentum is given by:
\eqn\mom{ P_\lambda = \sum^ N_{i=1} \left(\lambda_ i+(N-i)\beta \right)} 
This set of states is not complete, but 
the complete set can be obtained by multiplying these wave functions by $
\left(z_1z_2...z_N \right)^p $ 
where $ p $ is a positive or negative integer. Taking this into account, a 
complete basis  is given in terms of $ N $
momenta $ k_i=\lambda_ i+(N-i)\beta +p $ and the energy 
and momentum are given by:
\eqn\enmo{ \eqalign{ P \left\vert k_1...k_N \right\rangle &  = \left( \sum^
N_{i=1}k_i \right) \left\vert k_1...k_N \right\rangle   \cr H \left\vert
k_1...k_N \right\rangle &  = \left( \sum^ N_{i=1}k^2_i \right) \left\vert
k_1...k_N \right\rangle    \cr}} 
This looks like a free particle spectrum, 
in particular for  
$ \beta =0,1 $,  the Hamiltonian is a pure kinetic energy term  and it
reduces to 
the usual boson and fermion description of the states.
To understand the difference, let us consider the case of integer values of $
\beta . $ 
One sees from the expression of the $ k_i $ that they are integers and obey
the 
constraint:
\eqn\cont{ k_{i}-k_{i+1}\geq \beta }
If $ \beta =1, $ this is just the Pauli-exclusion
principle, but if $ \beta =2,3,... $ the Pauli 
principle is replaced by a stronger exclusion principle.
\vskip 15pt
\subsec {Particle hole excitations}

We shall now describe the excitations of the ground state of such a system 
and show that they can naturally be described in terms of quasiparticles. To 
illustrate this, we shall consider the matrix elements (form factor):
\eqn\form{ \langle \alpha \mid \rho (x)\mid 0\rangle }
of the density operator
\eqn\dens{ \rho (x)= \sum^ N_{i=1}\delta \left(\theta_ i-x \right) }
where the state $ \mid 0> $ stands for the ground state and $ \mid \alpha > $
for some excited 
state.
The results which we present here were motivated by the evaluation
of the density density correlation function obtained by Simons, Lee
and Altshuler in the matrix models \simon~(see also \hal).

The question we address is for which states $ \mid \alpha > $ is the matrix
element non 
zero?

In the case $ \beta =1 $ which corresponds to free fermions, it is easy
to answer 
using second-quantization arguments. The operator $ \rho (x) $ can be
represented 
as $ \psi^ +(x)\psi (x) $ where $ \psi (x) $ is a fermion field. Acting on
the vacuum, the 
operator $ \psi (x) $ anihilates
a particle from the Fermi sea $ (\mid k\mid \leq k_F $ where $ k_F
$ is the 
Fermi momentum) and the operator $ \psi^ +(x) $ creates a particle out of the
Fermi 
sea. The states $ \mid \alpha > $ are therefore described by $ (N-1) $ momenta
$ k_i $ less than the 
Fermi momentum $ k_F $ and 1 momentum $ k_N $ larger than $ k_F $ or
equivalently in terms 
of a hole of momentum less than $ k_F $ and a particle of momentum larger than
$ k_F. $

For $ \beta $ arbitrary integer,
the calculation is based on the properties of the wave 
functions of $ \tilde H $ 
(the Jack polynomials) established by Macdonald and Stanley
\mac \stan. The final 
result is however simple enough to be described here. Let us view the ground 
state as a Fermi-sea filled with $ N $ particles subject to the constraint $
\left\vert k_i-k_j \right\vert \geq \beta : $
\eqn\gro{ k_i=\beta \left(i-{N+1 \over 2} \right),\ \ \ \ 1\leq i\leq N  }
The Fermi momentum is given by the largest value of $ k_i: $
\eqn\ferm{ k_F=\beta{ N-1 \over 2}. }
This is represented on Figure 1 a) for $ \beta =2, $ $ N=5. $ The ones stand for
the 
integers $ k_i $ occupied by a particle and the 0 for the remaining integers.
\fig{The Fermi sea}{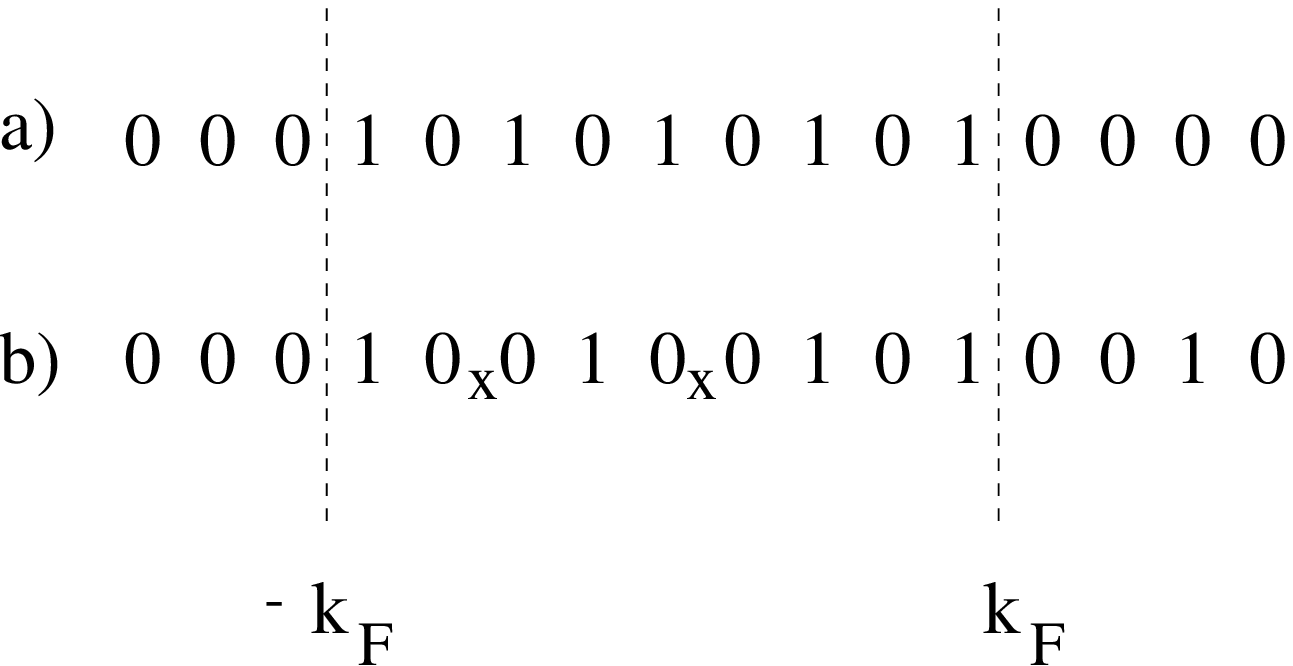}{8cm}

It turns out that the states $ \mid \alpha > $ which propagate are those for
which $ (N-1) $ 
momenta are inside the Fermi-sea $ \left\vert k_i \right\vert \leq k_F $ and
one momentum is out $ k_N>k_F $ (see 
Figure 1 b) ). Such states can be completely described by $ \beta $ holes of
momentum 
less than $ k_F $ and a particle of momentum larger than $ k_F. $ A hole
corresponds to 
a sequence of $ \beta $ consecutive integers not occupied by a momentum
(represented 
by a cross on the Figure).

The outcome of the computation \nouss ~
(These results have also been conjectured in \Haldconj~
and \polil)
is that these holes are eigenstates of a 
Calogero-Sutherland Hamiltonian with coupling constant $ \beta^{ -1} $ and
have a mass 
equal to $ {1 \over \beta} $ the mass of the particles.
In the thermodynamic limit the form factor is equal to~:
\eqn\introtherlim{
 \langle \alpha \mid \rho (0)\mid 0\rangle =
(w^2-1)^{{\beta-1}\over 2} \prod_{i=1}^\beta 
(1-v_i^2)^{{\beta^{-1}-1\over2}} 
 \prod_{i<j} \vert v_i-v_j \vert^{1/\beta}  
{(\Sigma_i v_i-\beta w) \over \prod_{i=1}^\beta (v_i-w) } }
where $v_i$ and $w$ are respectively the rapidities of the holes and
of the particle describing the intermediate state.
Other form factors such that $ \langle \lambda \mid \psi (x)\mid 0\rangle $
confirm this description of the 
excitations.
Although the description 
given here is reminiscent of a second quantized formalism, I am not aware 
that such a formalism exists.
\newsec {THE SPIN CALOGERO-SUTHERLAND MODELS}

A step towards the unification of the Calogero-Sutherland models and the 
integrable spin chains consists in introducing spin generalizations of the 
model studied in the last section.

The Hamiltonian we consider is
\eqn\hamspin{ H= \sum^ N_{j=1} \left(z_j{\partial \over \partial z_j} \right)^2- \sum^{
}_{ i\not= j}\beta \left(\beta -P_{ij} \right){z_iz_j \over \left(z_i-z_j
\right)^2} }
The particles have an internal spin degree of freedom and the permutation $
P_{ij} $ 
exchanges the spins of the particles $ i $ and $ j. $

The eigenstates have the following structure:
\eqn\eigen{ \psi \left(z_i,\sigma_ i \right)=\Phi \left(z_i,\sigma_ i \right) \prod^{
}_{ i<j} \left(z_i-z_j \right)^\beta }
where the wave function $ \phi \left(z_i,\sigma_ i \right) $ is completely
antisymmetric under the 
simultaneous permutations of the spins and coordinates.

For our purpose, it is more convenient to look at the effective Hamiltonian 
acting on $ \Phi \left(z_i,\sigma_ i \right): $
\eqn\efec{ \tilde H  = \sum^ N_{j=1} \left(z_i{\partial \over \partial
z_j} \right)^2+\beta \sum^{ }_{ i\not= j}{z_i+z_j \over z_i-z_j}
\left(z_i{\partial \over \partial z_i}-z_j{\partial \over \partial z_j}
\right)+\beta \sum^{ }_{ i\not= j} \left(P_{ij}+1 \right){z_iz_j \over
\left(z_i-z_j \right)^2  } } 
It turns out that this Hamiltonian can be diagonalized in a similar way as 
the one introduced in the last section. It has the same additive spectrum but 
the selection rules for the momenta differ from \cont.

To do this, let us define the following Dunkl operators:
\eqn\dun{ d_i=z_i{\partial \over \partial z_i}-\beta i-\beta \sum^{ }_{ j>i}{z_i
\over z_i-z_j} \left(K_{ij}-1 \right)+\beta \sum^{ }_{ j<i}{z_j \over z_j-z_i}
\left(K_{ij}-1 \right) }
where $ K_{ij} $ permutes the coordinates $ z_i $ and $ z_j: $
\eqn\truc{ \matrix{ \left[K_{ij},z_k \right]=0\ \ \ \ {\rm for} \ \ \ \ k\not= i,j
\hfill \cr K_{ij}z_j=z_iK_{ij} \hfill \cr}} 
Some motivation for the form of these operators and the algebra which
they obey will come later in the text
(see also \cher\poli), but first, let us see how they
can be used to solve the model.
We shall show that the diagonalization of $ H $ can be reduced to the 
simultaneous diagonalization of the Dunkl operators $ d_i. $ Moreover in a 
convenient basis the $ d_i $ are represented by triangular matrices.

The $ d_i $ obey the defining relations of a degenerate affine-Hecke algebra:
\eqn\afe{ \eqalign{ a) & \ \ \ \left[d_i,d_j \right]=0   \cr b) & \ \ \
\left[K_{ii+1},d_k \right]=0\ \ \ {\rm if} \ \ \ k\not= i,i+1   \cr c) & \ \ \
K_{i,i+1}d_i-d_{i+1}K_{i,i+1}+\beta =0   \cr} }
From these relations, one deduce that the quantity:
\eqn\du{ D(u)= \prod^ N_{i=1} \left(u-d_i \right)=u^N+ \sum^ N_{k=1}c_ku^{N-k} 
}
commutes with the permutations $ K_{ij}. $

By computing explicitely $ \sum^{ }_{ } d^2_i=c^2_1-2c_2, $ one sees that it
is equal to the 
Hamiltonian $ \tilde H $ except that the $ P_{ij} $ are replaced by $ -K_{ij}.
$ Since $ D(u) $ commutes 
with the permutations, we can restrict to the space of antisymmetric wave 
functions $ \phi \left(z_i,\sigma_ i \right). $ In this subspace $ -K_{ij} $
and $ P_{ij} $ coincide and $ \sum^{ }_{ } d^2_i $ is equal 
to $ \tilde H. $ Thus, instead of $ \tilde H, $ it is easier to simultanesouly
diagonalize the $ d_i $
's.

If we consider the explicit expression \dun
~of the $ d_i $'s, one sees that 
they preserve the space of homogeneous polynomials of a given degree in the 
variables $ z_i. $ It then follows from simple arguments 
that these 
operators are realized by triangular matrices in the monomial basis $
z^\lambda =z^{\lambda_ 1}_1...z^{\lambda_ n}_n $ \nous.
Their eigenvalues are equal (up to a permutation) to the components of the 
multiplet:
\eqn\multip{ \delta_ i=\lambda_ i+\beta (N-i)}
The eigenvectors of $ \sum^{ }_{ } d^2_i $ corresponding to a given eigenvalue
form a 
representation of the permutation algebra. The wave functions $ \phi
\left(z_i,\sigma_ i \right) $ are 
obtained by antisymmetrizing these eigenvectors with an arbitrary spin 
function $ X\left(\sigma_ i \right). $ In particular, if we antisymmetrize
the wave functions, we 
recover the Jack polynomials of the last section (at coupling $ \beta +1) $
multiplied 
by the Vandermonde determinant $ \prod^{ }_{ i<j} \left(z_i-z_j \right). $
The mathematical 
properties of the Jack polynomials can probably 
be extanded to these wave-functions.
\vskip 15pt
\subsec {Dunkl operators and monodromy matrices}

It is clear from the explicit construction of the wave functions of $ H $ that
it's spectrum is highly degenerate and we shall now interpret these 
degeneracies from a symmetry principle. For this, we need to introduce the 
so-called monodromy matrix $ T(u) $ solution of the equation:
\eqn\mono{ R^{ab}T^aT^b=T^bT^aR^{ab}} 
where $ T^a $ in the operator valued matrix $ T \left(u_a \right)\otimes 1, $
$ T^b $ is $ 1\otimes T \left(u_b \right) $ and $ R^{ab} $ is 
given by
\eqn\rmat{ R^{ab}=u_a-u_b+\beta P_{ab} }
where $ P^{ab} $ permutes the two components of the tensor product $
V^a\otimes V^b. $

Consistency of the above relations requires that $ R^{ab} $ is a solution of
the 
Yang-Baxter equation in $ V^a\otimes V^b\otimes V^c: $
\eqn\bax { R^{ab}R^{ac}R^{bc}=R^{bc}R^{ac}R^{ab} }
Assume that $ T(u) $ has a series expansion in $ 1/u, $ the coefficients of
this 
expansion define an algebra which is called the Yangian \drin.
In this 
section, we exhibit a representation of the Yangian which commutes with $
\tilde H. $ 
For this, let us define the so-called \lq\lq$ L $ operators \rq\rq \fad~
defined as :
\eqn\lop{  L_i=u-d_i-\beta P_{ai} } 
where $ P^{ai} $ permutes the components of the tensor product $ V^a\otimes
V^i $ and $ d_i $ is a 
coefficient which we identify with the Dunkl operator \dun.
This is consistent since the $d_i$'s commute among themselves and with the
permutations $P_{ij}$.
A well known representation of the monodromy matrix $ T(u) $
is 
then given by \fad:
\eqn\mona { T^a(u)=L_1(u)L_2(u)...L_N(u) } 
Now, let us act with this monodromy matrix on the space of wave functions $
\phi \left(z_i,\sigma_ i \right). $ 
The coordinates $ z_i $ are permuted by the $ K_{ij} $ and the spins $ \sigma_
i $ are permuted by 
the $ P_{ij}. $ We shall see that the antisymmetric wave functions $ \phi
\left(z_i,\sigma_ i \right) $ are 
preserved by the action of the monodromy matrix $ T(u). $ To show this, we
define 
a projection $ \Pi $ which substitutes $ -P_{ij} $ for $ K_{ij} $ to the right
of an expression.

The subspace of antisymmetric wave functions is preserved if the following 
condition is satisfied:
\eqn\cond{ \Pi \left( \left(K_{ij}+P_{ij} \right)T(u) \right)=0 }
which is equivalent to the condition:
\eqn\eqi { \Pi \left( \left(K_{ii+1}+P_{ii+1} \right)L_i(u)L_{i+1}(u) \right)=0}
when we expand this relation in powers of $ u, $ the coefficients of $ u $ and
$ u^2 $ 
give relations which are trivially satisfied. Let us here consider the 
coefficient of 1. We omit the $ \Pi $ in front of the expression:
\eqn\glo{
0= \left(K_{12}+P_{12} \right) \left(d_1d_2-\beta d_1P_{a2}-\beta
d_2P_{a1}+\beta^ 2P_{a1}P_{a2} \right) }
using \afe~and the fact that $
\left[K_{12},d_1d_2 \right]=0 $ we obtain:
\eqn\glor{ \eqalign{ -\beta P_{a1}P_{a2}P_{12}+\beta P_{12}P_{a1}P_{a2} &\cr
-d_2P_{12}P_{a1}-d_1P_{12}P_{a2} &\cr + P_{a1} \left(d_1P_{12}-\beta
\right)+P_{a2} \left(d_2P_{12}+\beta \right) & =0 \cr} }
which is easily shown. Notice
that 
the unexpected term equal to $ \beta $ in \afe c)
is necessary for the proof of this 
relation.

It follows from this analysis that the monodromy matrix $ T(u) $ generates a 
representation of the Yangian algebra in the subspace of antisymmetric wave 
functions $ \phi \left(z_i,\sigma_ i \right). $ It is also easy to see that
this representation commutes 
with the Hamiltonian $ \tilde H $ and therefore, the Yangian is the symmetry
algebra 
which explains the degeneracies of $ \tilde H. $
\vskip 15pt
\subsec{Yang's representation of affine-Hecke algebras}

In this section, we establish a relation between the Dunkl operators
\dun~
and some operators defined by Yang in his study of the $ \delta $ interacting
gas (see also \fel~for a discussion in the elliptic case). 
%This relation is also considered by 
%V.M.Buchstaber,G.Felder and A.P.Veselov in the more general 
%case of the elliptic Dunkl operators.

To make this relation more transparent, we shall consider the case of the 
affine-Hecke algebra defined by :
\eqn\afel{ \eqalign{ a) & \ \ \ \left[y_i,y_j \right]=0 \cr b) & \ \ \
\left[g_i,y_j \right]=0\ \ \ {\rm if} \ \ \ j\not= i,i+1  \cr c) & \ \ \
g_iy_i-y_{i+1}g^{-1}_i=0 \cr} }
and the $ g_i $ obey the relations of a Hecke algebra :
\eqn\hec{ \eqalign{ &\left[g_i,g_j \right]  =0\ \ \ {\rm if} \ \ \ \vert i-j\vert
\geq 2   \cr &g_ig_{i+1}g_i  =g_{i+1}g_ig_{i+1}   \cr &\left(g_i-q \right)
\left(g_i+q^{-1} \right)  =0  \cr} }
This algebra replaces the permutations and the $ d_i $'s  if one repeats
the 
arguments of the last section 
replacing the rational solution \rmat~ of the Yang-Baxter equation \bax~
with a trigonometric solution. 
This is considered in \nous.
This algebra is also studied in the works of
I.Cherednick \cher.
Here we shall content ourselves to obtain a natural 
representation of the affine-Hecke algebra \afel~
which in some limit degenerates to the 
Dunkl operators \dun.

For this, let us consider some realization of the Hecke-algebra \hec~
acting in the tensor product $ V\otimes^ N $ where the operator $ g_i $ acts
in $ V_i\otimes V_{i+1}. $ To 
ease the forthcoming discussion, we rename $ g_i: g_{i,i+1}. $
Following Yang's argument \gaud , we shall exhibit a set of operators which commute 
together. Then we shall see that in some limit these operators can be 
identified with the $ y_j. $

The starting point is the trigonometric form of the $ L $
operator \fad ~
given by:
\eqn\newl {L^a_i(v)= \left({v \over v_i}g_{ia}-g^{-1}_{ia} \right)K_{ai}}
where $g_{ai}$ acts in $ V_a\otimes V_i$ and $ K_{ai} $ permutes the two components of $ V_a\otimes V_i. $
Using this form we can construct a monodromy matrix 
solution of \mono~ given by :
\eqn\newmo{ T^a(u)=L_1(u)L_2(u)...L_N(u)S^a }
where $S^a$ is a matrix acting in $V^a$ and such that $[R^{ab},S^a \otimes S^b]=0$.
It then 
follows from the Yang-Baxter equation \mono~  that the trace of this monodromy matrix 
$ \bar T(v) $ defines a commuting set of transfer matrices:
\eqn\comu{ \left[\bar T(v),\bar T(w) \right]=0\ \ \ \ \ \ \forall u ,\ \forall v }

To define the trace of the monodromy matrix, we consider $ T^a $ as a matrix 
acting in $ V^a $ and take its trace in the usual sense.

Hence, the
operators $ \tilde y_i=\bar T \left(v_i \right) $ form a commuting 
set . These operators play an important role in the study of the
$ \delta $ interacting gas \gaud.
They can be explicitely computed using the fact 
that:
\eqn\tro{ L^a_i \left(v_i \right)=\left(q-q^{-1} \right)K_{ai}
}
If one then  writes $ \bar T \left(v_i \right) $ in the form $ \bar T
\left(v_i \right)= {\rm tr} \ K_{ia}\theta $ where $ \theta $ is an 
operator that contains only the permutations $ K_{ij} $ with $ i,j\not= a $
and uses the 
following property of the trace:
\eqn\tra{ {\rm tr} \ K_{ai}\theta =\theta}
one finally obtains :
\eqn\tilda{ \tilde y_i=X_{ii+1}X_{ii+2}...X_{iN}S^iX_{i1}...X_{ii-1} }
where the operators $ X_{ij} $ are given by:
\eqn\xij{ X_{ij}= \left({v_i \over v_j}g_{ij}-g^{-1}_{ij} \right)K_{ij} }
The operators $ \tilde y_i $ constructed in this way obey the relation
\afel a) but not 
\afel b) c). 
It is then not difficult to see that these relations 
are satisfied in the limit $ 0\ll v_1\ll v_2...\ll v_N. $ 
In this limit  the operators $ y_j $ are given by \nous \cherll :
\eqn\fina{
y_i=g^{-1}_{ii+1}K_{ii+1}...g^{-1}_{iN}K_{iN}S^iK_{1i}g_{1i}...K_{i-1,i}g_{i-1,i}
}

To make contact with the Dunkl operators $d_i$, we must use the following
representation of the Hecke algebra acting in the space of polynomials in
N variables $z_i$ :
\eqn\repg{g_{ii+1}=qK_{ii+1}-(q-q^{-1}){z_i\over z_i-z_{i+1}}(K_{ii+1}-1) }
and $S^i$ is defined by:
\eqn\si {(S^i f)(z_1,...,z_N)=f(z_1,...,tz_i,...,z_N) }
If one sets $q=t^{\beta}$ and let $t\to 1$, the operators $y_i$ tend to
the $d_i$ defined in \dun.

\newsec{THE HALDANE SHASTRY CHAIN}
Among this class of models, the Haldane Shastry chain  \ha\sha~is the closest to the
known integrable spin chains such as the XXX chain \gaud.
It is also the less well understood.
Here, we shall present these chains from the point of view of the Dunkl
operators.

The Hamiltonian is the $\beta=\infty$ limit of \hamspin, it is equal to :
\eqn\hasha{
H= \sum_{i<j} {P_{ij}\over \sin^2 (\theta_i -\theta_j)} }
where the variables $\theta_j$ take the values ${\pi j\over N},\  0\leq j<N $.
The main difficulty comes from the fact that this Hamiltonian,
unlike \hamspin, cannot be obtained from the static limit of the Dunkl
operators as $\sum d_i^2$ because in this limit this quantity is a number.
In this section, we define a representation of the degenerate affine
Hecke algebra \afe~labelled by N complex numbers $\omega_k$.
When the $\omega_k$ are arbitrary, the representation is irreducible,
when they are equal to $\exp(i{2\pi k\over N})$, it is reducible and one can find the
Haldane Shastry Hamiltonian in its center.
We shall then use a correspondence with the $\beta=1$ representation
\dun~ to diagonalise the Hamiltonian \hasha.

Consider the representation of the degenerate affine Hecke algebra \afe
~given by:
\eqn\dunl{ \tilde d_i=\sum_{ j>i}{z_i
\over z_i-z_j} K_{ij}- \sum_{ j<i}{z_j \over z_j-z_i}
K_{ij} }
It acts on the Hilbert space defined by the permutations :
\eqn\hilb{
{\cal H}_N = \{ \vert P_1...P_N> / P_i\in Z_N ,\ P_i\ne  P_j \ \ \ \rm{for\ all} \ i,j \} }
in the following way :
\eqn\folo{
\eqalign{
K_{ij}\vert ...P_i...P_j...>\  &=\ \vert ...P_j...P_i...> \cr
z_j \vert P_1...P_N>\  &=\ \omega_{P_j}\vert P_1...P_N> \cr  }}
In the case where the $\omega_k$'s are arbitrary, there is no reason for this
representation to be reducible. On the other hand, if $\omega_k=
e^{i{2\pi k\over N}}$, the operator $C$ defined by:
\eqn\opc{ C\vert P_j>\ = \vert (P_j+1)> }
obviously commutes with the $d_j$'s and the $K_{ij}$.
So, in this case the representation is reducible and one can verify 
that the Hamiltonian given by :
\eqn\hamar{\tilde H= \sum _{i<j} {z_iz_j \over (z_i-z_j)^2 }K_{ij}}
is also in the center. No systematic way to obtain the operators which
commute with the $K_{ij}$'s and the $d_i$'s is known.
In what follows we restrict to the case where 
$\omega _k=\exp (i{2\pi k\over N})$.

The Hamiltonian \hamar~ can easilly be related to the Haldane Shastry
Hamiltonian and we shall now indicate how to obtain its eigevalues
and some of its eigenvectors. We shall construct  
a class of states on which the two representations
of the $d_i$ \dun~ (for $\beta=1$) and \dunl~ act in the same way.
On these states, any conserved quantity of the spin chain has its
analogue in the dynamical model and can thus be diagonalised .

For $M<N$ let us define the Hilbert space ${\cal H}_M$
\eqn\calh{
{\cal H}_M = \{ \vert P_1...P_M> / P_i\in Z_N ,\ P_i\ne  P_j \ \ \ \rm{for\ all} \ i,j \} }
which is uniquely embedded in ${\cal H}_N$ if one requires that :
\eqn\cocal{
K_{ij} \vert P_1...P_M> \ =\ 
 \vert P_1...P_M> \ \ \ \rm{for}\ i,j> M}
In ${\cal H}_M$ we consider the following states :
\eqn\states{
\vert \alpha>\ =\ \Psi (\omega^{P_1},...,\omega^{P_M})\vert P_1,...,P_M>  }
where :
\eqn\psi{
\Psi(z_i)\ =\ \Phi(z_i) \prod _{i<j} (z_i-z_j) }
and $\Psi(z_i)$ is a polynomial of degree between 1 and $N-1$ in
the variables $z_i$.
On these states, 
$\tilde d_i=d_i-{N+1\over2}$ for $1\le i\le M$.
It results from the fact that a polynomial $P(z)$
of degree between $1$ and $N-1$ satisfies the relation :
\eqn\rela{
(z\partial_z - {N+1\over 2}) P(z)\vert_{z=\omega_k}
=\sum_{l\ne k} {\omega_k\over \omega_k-\omega_l}P(\omega_l) }
The same relation enables to show that in this subspace $\tilde H$ acts 
as $\sum_{k=1}^M (d_k +1/2+N/2)(d_k +1/2-N/2)$ and to  find its
spectrum:
\eqn\spect{
E=\sum_{k=1}^M\epsilon(f_k)  \ \ \ \  \rm{with}  
\ \ \epsilon(f_k)=(f_k+1/2)^2-N^2/4 }
$f_k+{N+1\over2}$ are the degrees of the highest monomial of $\Psi(z_i)$.
Therefore, the $f_k$ satisfy the inequalities:
\eqn\ineq{
-{N-1\over 2}\leq f_1 \leq f_2 ...\leq {N-1\over 2}-1 }

\listrefs

\end